\documentstyle[11pt,aaspp4]{article}
\def\beq{\begin{equation}}
\def\eeq{\end{equation}}
\def\bey{\begin{eqnarray}}
\def\eey{\end{eqnarray}}
\def\kms{\mbox{\rm \,km\,s}^{-1}}
\def\msun{M_\odot}
\def\mag{~mag}
\def\etal{et al.}
\input epsf

\begin{document}

\title
	{Magnitude bias of microlensed sources towards the Large Magellanic Cloud}
\author
	{HongSheng Zhao
	\\Sterrewacht Leiden, 
Niels Bohrweg 2, 2333 CA, Leiden, The Netherlands (hsz@strw.LeidenUniv.nl)}
\author
	{David S. Graff
	\\Departments of Physics and Astronomy,
The Ohio State University, Columbus, OH 43210, USA (graff.25@osu.edu)}	
\author
	{Puragra Guhathakurta\altaffilmark{1}
	\\UCO/Lick Observatory, University of California at Santa Cruz,
Santa Cruz, CA 95064, USA (raja@ucolick.org)}
\altaffiltext{1}{Alfred P. Sloan Research Fellow}
\date{Accepted ........      Received .......;      in original form .......}
\label{firstpage}

\begin{abstract}
There are lines of evidence suggesting that some of the observed
microlensing events in the direction of the Large Magellanic Cloud
(LMC) are caused by ordinary star lenses as opposed to dark Machos in
the Galactic halo.  Efficient lensing by ordinary stars generally
requires the presence of one or more additional concentrations of
stars along the line of sight to the LMC disk.  If such a population
behind the LMC disk exists, then the source stars (for lensing by LMC
disk objects) will be drawn preferentially from the background
population and will show systematic differences from LMC field stars.
One such difference is that the (lensed) source stars will be farther
away than the average LMC field stars, and this should be reflected in
their apparent baseline magnitudes.  We focus on red clump stars:
these should appear in the color-magnitude diagram at a few tenths of
a magnitude fainter than the field red clump.  Suggestively, one of
the two near-clump confirmed events, MACHO-LMC-1, is a few tenths of
magnitude fainter than the clump.
\end{abstract}


\section{Introduction} 

One way of hiding most of the baryons in the Universe is to lock them
in Machos, baryonic dark matter in the form of compact objects in
galaxy halos.  This has motivated current experimental searches for these
objects in the Milky Way halo via gravitational microlensing
(Pacz\'nyski 1986), the temporary brightening of one of out of a
million background stars as the unseen Macho passes close to the line
of sight of that star by chance and focuses its light gravitationally.
About $20-30$ microlensing events in our line of sight to the Large Magellanic
Cloud (LMC) and two to the Small Magellanic Cloud have been detected
by the MACHO and EROS groups (Alcock \etal\ 1997a; Renault \etal\ 1997;
Alcock \etal\ 1997b; Palanque-Delabrouille 1998;
Alcock \etal\ 1999; Alcock \etal\ 2000), including two caustic binary events
(MACHO-98-SMC-1 and MACHO-LMC-9) which are unambiguous cases of
microlensing events.  Nevertheless, it is a subject of heated debate
whether these microlensing events are indeed due to dark matter in the
Galaxy's halo.

Currently there are two popular views on the issue:
\begin{itemize}
\item Galactic Halo Lensing Hypothesis (Machos): The lenses are
located in the halo, and are dark.  These lenses could be baryonic
dark matter candidates (Alcock et al. 1997a, 2000; Gates \& Gyuk 1999);
\item Lensing by Ordinary Stars Hypothesis: The lenses
are ordinary stars, not dark matter.  This hypothesis is strongly
constrained so that the only presently viable model is that the LMC
sports some extra three-dimensional thick stellar structure displaced from
the two-dimensional thin and cold disk of the LMC.  These stars are
providing either lenses or sources of the observed events (Zhao
1998a,b, 1999a).
\end{itemize}

A number of papers, including a few recent ones (Evans \& Kerins 1999,
Graff \etal\ 1999a; Fields, Freese \& Graff, 1999; Gates \& Gyuk 1999, Gyuk, Dalal \& Griest 1999,
Zhao 1999a) have investigated the shortcomings and plausibility of the
two scenarios.  Here we suggest using the source magnitude
distribution to differentiate between the two classes of models.  The
existing data are not conclusive; however, there should shortly be
enough data to apply the method discussed here.
 
Ordinary star lensing models require additional concentrations of
stars along the line of sight to the LMC, in front of its main disk,
behind it, or perhaps both (e.g.,~a partial or complete ring around
the LMC).  The added population may either be an extended distribution of
stars around the LMC (e.g.,~tidal extension) or a distinct stellar
system.  If additional stars are located behind the LMC disk, then the
sample of {\it source stars\/} that undergo lensing by ordinary stars
in the LMC disk will be strongly biased to be in the background
population.  On the other hand, if lensing is due to Galactic halo
Machos or a foreground object, the sample of lensed stars should be a
random subsample of the observed stars in the LMC, weighted only by
the microlensing efficiency of observing an event in that particular
star.  The observational consequences of the lensed stars being behind the LMC
disk are that they should have:
\begin{itemize}
\item Slightly fainter baseline magnitudes in all bands than unlensed stars
due to a larger (by $\sim 0.3$\mag) distance modulus;
\item Preferentially fainter apparent magnitudes in the bluer bands than
unlensed stars in neighbouring lines of sight due to reddening by intervening
dust in the LMC disk: $\sim 0.6$\mag of
$U$-band extinction with factor of two variations on arcminute angular scales
(cf.~Zhao 1999c,d); and
\item Velocity offsets of up to $\sim 20$~km~s$^{-1}$ relative to unlensed
stars because the kinematics of background stars will, in general, be
different from those in the LMC disk (Zhao 1999b; Graff \etal\ 1999b).
\item Small lens-source proper motions ($\sim 30$km/s at the LMC/SMC)
measurable with caustic crossing events (Kerins \& Evans 1999).
\end{itemize}

As reviewed by Zhao (1999a), all four effects statistically
differentiate between the LMC self lensing and Macho hypotheses and do
not necessarily apply on a star-by-star basis.  A nice feature of
these effects is that they are observable not only in real time for an
on-going microlensing event but also for long-past events; the latter
allows flexible telescope scheduling.  These effects do {\it not\/}
occur in the Galactic halo lensing models since the stars in the back
and front of the LMC have nearly equal chance of being lensed by a
Macho halfway to the LMC.  In this paper, we examine the bias in the
apparent baseline magnitudes of microlensed stars (the magnitude
observed in the star either before or after the lensing event).  This
effect has been first discussed by Stanek (1995) for microlensing
events in the Galactic bulge.

\section{Model of the source and lens distribution}

Based on microlensing experiments alone, it is possible to constrain the
positions of a putative additional object along the line of sight to
the LMC.  Objects bound to the Magellanic clouds have a typical
transverse velocity of $70 \kms$ (the speed of rotation of LMC disk
stars) relative to the systemic velocity.  Assuming a typical lens mass
of $0.1-0.3 \msun$ from a standard stellar IMF, and a typical Einstein
radius crossing time of 45 days, then the lens-source distance is of
order $5-10$~kpc.
The detailed models of Zhao (1999b)  suggest similar distances.  The
models of Evans \& Kerins (1999) and the N-body simulation of
Weinberg (1999) have a wide range of source-lens distances with a mean
value also in the range $5-12.5$~kpc.

We assume that stars in the direction of the LMC are distributed in
two separate groups, the primary LMC disk having a surface mass
density $\Sigma_{\rm D}$ and an extra background population with
surface density $\Sigma_{\rm B}$ situated $5-12.5$~kpc (0.2 -- 0.5
mag) behind the primary disk, and we have tried a number of distance
distributions within this range.  We define $\Sigma_{\rm tot} \equiv
\Sigma_{\rm D}+\Sigma_{\rm B}$.  There may also be standard Galactic
halo Machos between us and the LMC.  The implications of a possible
population in the immediate foreground of the LMC disk will be
discussed later.

We define $f_{\rm B}$ to be the fraction of lensing events in which
the lensed (source) stars are behind the LMC disk.  Thus, $f_{\rm
B}=0$ is for all halo lensing models in which all the lenses are well
in front of the LMC and all the lensed (source) stars are in the LMC
disk.  Even in models in which there are additional stars behind the
LMC disk, we expect $f_{\rm B} < 0.9$ because some of the lensing
events will be due to self lensing within the LMC disk (Wu 1994; Sahu
1994; Gould 1995).

\section{The luminosity function of unlensed and lensed clumps}

In this paper, we investigate small ($\sim 0.3$\mag) differences in
apparent baseline magnitude between lensed stars and field stars.  The
only place in the color-magnitude diagram where there is a feature sharp 
enough for this difference to
be seen and which contains enough stars to be important is the red clump.
(cf. Figure~1 of astro-ph/9907348).
The red clump is a sharp feature, with width of only 0.14\mag in the $I$
band.  It has been used as a standard candle on several occasions in the
past; in particular, Stanek (1995) applied it to microlensing events 
in the Galactic bulge.

  Even though main sequence events outnumber
clump events by a factor of $~10$, main sequence events cannot be used
for this analysis since the main sequence is nearly verticle.  A small
shift in magnitude of a main sequence star will leave that star in the
main sequence, especially when we consider that background stars are
also likely to be reddened (Zhao 1999c)

Our method is applicable to any passband, and is presented here for a
general passband $X$.  We suggest that microlensing groups apply the
method to the clump luminosity function in their particular passbands.
We advocate the $I$ band because the clump is narrow in $I$
(cf. Figure~1).

We parametrize the observed apparent $X$-band luminosity function of the clump 
in the LMC disk
as a narrow Gaussian superposed on a broader Schechter-like luminosity
function for red giants.
\beq\label{lf}
n_{\rm D}(X)=
 C_1 \exp\left[-0.5\left({X-X^{\rm RC} \over \sigma^{\rm RC}}\right)^2\right]
+C_2 F^\alpha \exp\left(-{\beta F \over F^{\rm RC}}\right),
\eeq
where $X-X^{\rm RC}$ and $F/F^{\rm RC}$ are the observed $X$-band magnitude and flux of a 
star {\it relative} to that of an average red clump, 
$\sigma^{\rm RC}$ is the width (dispersion) of the observed clump distribution,
$C_1$, $C_2$, $\alpha$ and $\beta$ are constants to be adjusted to fit the observed
luminosity function.  Assuming that the stellar populations in
the LMC disk $(D)$ and background object $(B)$
are the same, the magnitude distribution of the background
objects $n_{\rm B}(X)$ is simply that of the disk stars 
$n_{\rm D}(X)$ shifted by some amount $\Delta$
towards the faint side due to distance modulus and any excess extinction,
i.e.,
\beq
n_{\rm B}(X)= 
{\int_{{\rm Min}(\Delta)}^{{\rm Max}(\Delta)} d\Delta P(\Delta) n_{\rm D}(X-\Delta) 
\over
\int_{{\rm Min}(\Delta)}^{{\rm Max}(\Delta)} d\Delta P(\Delta) },
\eeq 
where $P(\Delta)$ is the probability of finding 
a background population with a magnitude 
shift of $\Delta$.  
Assume equal chance of finding the background population
at a distance $(1.1-1.25)\times 50$ kpc, we have
\beq
P(\Delta) \approx {\rm const},~~~
{\rm Min}(\Delta) \approx 0.2~{\rm mag},~~~{\rm Max}(\Delta) \approx 0.5~{\rm mag}.
\eeq

The observed luminosity function of LMC stars is a superposition
of the disk stars and the background stars.  
However, in the limiting case that most of the surface density is
in the LMC disk, we can make the simplifying assumption 
\beq
n_{\rm obs}(X)=n_{\rm D}(X)+ \frac{\Sigma_{\rm B}}{\Sigma_{\rm D}} n_{\rm B}(X) \approx n_{\rm D}(X).
\eeq
Fitting eq.~(\ref{lf}) directly to observations, here the published 
OGLE observed clump magnitude distribution in the $I$ band (Udalski \etal\ 1998), 
we determine the fitting parameters $C_1$, $C_2$, $\alpha$ and $\beta$; 
the dispersion of the clump $\sigma^{\rm RC}=0.14$\mag is set at the value 
given by Udalski \etal.
The lower panel of Figure~1 shows our model of the observed luminosity function
is a fair approximation to that observed by the 
OGLE survey (cf.~Fig.~5 of Udalski \etal\ 1998).
Working with the relative magnitudes $X-X^{\rm RC}$ has the advantage that 
the distributions are independent of calibrations of the zero point of the clump,
and insensitive to changes of the passbands.
Note that the derived luminosity function are {\it not} corrected for reddening, and 
we will consistently use uncorrected magnitudes throughout the paper.

The luminosity function of the lensed stars depends on the fraction of
lensed stars in the LMC disk and the fraction of lensed stars behind
the LMC:
\beq
n_{\rm source}(X,f_{\rm B})=f_{\rm F} n_{\rm D}(X) + f_{\rm B} n_{\rm B}(X).
\eeq
Here $f_{\rm F} \equiv (1-f_{\rm B})$ is the fraction of events due to foreground lenses
and LMC thin disk sources, while 
$f_{\rm B}$ is the fraction of events due to LMC thin disk lenses
and background sources.  

If most of the lensing is due to a background object and $f_{\rm B}$ is
large, then the luminosity function of lensed stars will basically be
the luminosity function of unlensed stars shifted by $\Delta \approx 
(0.2-0.5)$\mag.  This shift is quite significant
since it is larger than the width of the clump, 0.14\mag in OGLE $I$ band.
As shown in the lower panel of Figure~1, the expected
lensed star luminosity function for a model with $f_{\rm B}=0.9$ is quite
distinct from the unlensed luminosity function.

\section{Analysis}

\subsection{Observed clump events}

  We urge the reader to examine Figure 3 of Alcock \etal\ (2000) for
this discussion.

  At the moment, there are at least $3$ confirmed events in the clump
region.  The latest MACHO collaboration publication (Alcock \etal\
2000) shows that events LMC-MACHO-1 is indeed a few tenths magnitudes
dimmer than the clump.  Although one event cannot by itself be
significant, this event is a few tenths of a mag fainter than the
clump, exactly where we expect it to be if lensing is due to a
background object.  LMC-MACHO-16 is brighter than the clump but this
event carries no statistical significance since it must be a red giant
branch object whether or not it is in the LMC disk or behind the disk.
LMC-MACHO-25 is in the center of the clump, and may be due to an LMC
disk source, indicating that $f_B <1$.  However, these few events are
not enough to make any significant statements about $f_B$.

These three events are shown in the upper panel of Figure~1, where the probability of finding a clump is 
a factor of a few lower than average.  The histograms here were made simply by binning
the MACHO LMC color-magnitude diagram in the MACHO-defined clump region $0.3 \le V-R \le 0.9$ and 
$17.5 \le V \le 20$ (cf. the clump ``square'' in Figure~11 of Alcock \etal\ 1997a).
These distributions we get are similar to the published OGLE $I$ magnitude
distributions (cf. Figure~5 of Udalski et al. 1998 and 
the lower panel of our Figure~2), but the OGLE distributions are narrower.
This is perhaps partly due to better seeing with OGLE and
a narrower color range $0.8<V-I<0.95$ for the clump in Udalski \etal\ (1998).
Furthermore, the clump is approximately horizontal in $I$ (Paczy\'nski \& Stanek
1998)---i.e.,~the mean magnitude of the clump does not depend strongly
on color.  Differential extinction in the $I$-band is also smaller,
making the clump a narrower feature in $I$ than in $V$ and $R$.
For these reasons we prefer to do our analysis in the $I$ band.
Unfortunately the offset $I$ magnitude of the MACHO-LMC-1 from the
clump is unknown, although it is perhaps fair to assume it is not too different
from the observed value in $V$ and $R$.
In this case the event lies also well off the peak for the unlensed stars, but
exactly under the peak of the luminosity function of lensed stars in the $f_{\rm B}=0.9$
model, a model in which most of the lensing is due to a background object.
The probability of finding such an event is approximately 5 times
greater in the $f_{\rm B}=0.9$ model than in the $f_{\rm B}=0$ model.  Thus,
formally, this single event favors the background lensing hypothesis
at the statistically marginal 80\% confidence level.

\subsection{Future clump events}

We have run Monte Carlo simulations to estimate the effect of increasing
the events.  These simulations are done in the $I$ band in order to
use the published OGLE clump luminosity function (Udalski et al. 1998).
These simulations suggest that perhaps 10
additional clump event is needed to exclude the Galactic halo lensing hypothesis.  

Extrapolating the present detection rate,
we anticipate that these clump events could be detected with
the OGLE~II and EROS~II experiments in the next few years.  
It should be very interesting to re-apply the above analysis.  

Finally we remark that in case Occam's razor fails, i.e.,~in case the dark halo is a mix of
Machos and non-baryonic matter and the microlenses are a mix
of Machos and stars (in the foreground or the LMC), then $f_{\rm B}$ can still
set an {\it upper limit\/} on the fraction of Machos in the halo via the
relation:
\beq
f_{\rm Macho} \le 0.2f_{\rm F} =0.2(1-f_{\rm B}),
\eeq
where the factor $0.2$ is the current estimated Macho halo mass fraction
(Alcock \etal\ 2000).

\section{Conclusion}

If the LMC lensing is due to the sources being in a background stellar
system, then the properties of lensed stars should be systematically
different from those of the bulk of LMC stars.  Although there are
several obvious comparisons that can be made (e.g.,~of the kinematics), the
simplest way of examining the lensed stars is to compare their
baseline apparent magnitudes with those of unlensed LMC field stars.
In particular, the red clump provides a sharp feature that is ideal
for such a comparison.

This technique is powerful because the expected distance difference modulus we
are searching for ($0.2 - 0.5$\mag) is wider than the sharp (0.14\mag) red
clump.  Any signal found using this (differential) technique cannot be due to
the normal spatially-variable reddening in the LMC disk, poor photometry, or
blending in crowded fields, all of which
affect lensed and unlensed stars in the LMC disk {\it equally}.

The technique can also apply to an arc-like distribution wrapping
around the LMC disk (Kunkel \etal\ 1997), with some stars in the
foreground and some in the background or to a shroud of stellar matter
around the LMC (Weinberg 1999, Evans \& Kerins 1999).  We note,
however, that this technique (and in fact all first three techniques
mentioned in the Introduction) cannot distinguish between the standard
Galactic halo lens model and a model in which there is a small
additional distribution of stars immediately in front of the LMC (but
none in the background), since in both cases the lensed stars will be
exactly at the distance of the LMC primary disk.

As few as one additional clump event could potentially rule out the
halo lensing hypothesis at the 95\% confidence level, although more
events will be needed if, as seems indicated by unpublished
microlensing alerts, some of the lensing events are due to a
foreground object.  Many more clump events should be available in a
few years, and these should yield valuable clues to the line of sight
structure of the Magellanic Cloud system.

We thank the referee, David Bennett for doing a very careful job in
scrutinization of the paper, David Alves, Andy Gould and Tim de Zeeuw for
helpful discussions.

{}

\begin{figure}
\epsfysize=15cm \centerline{\epsfbox{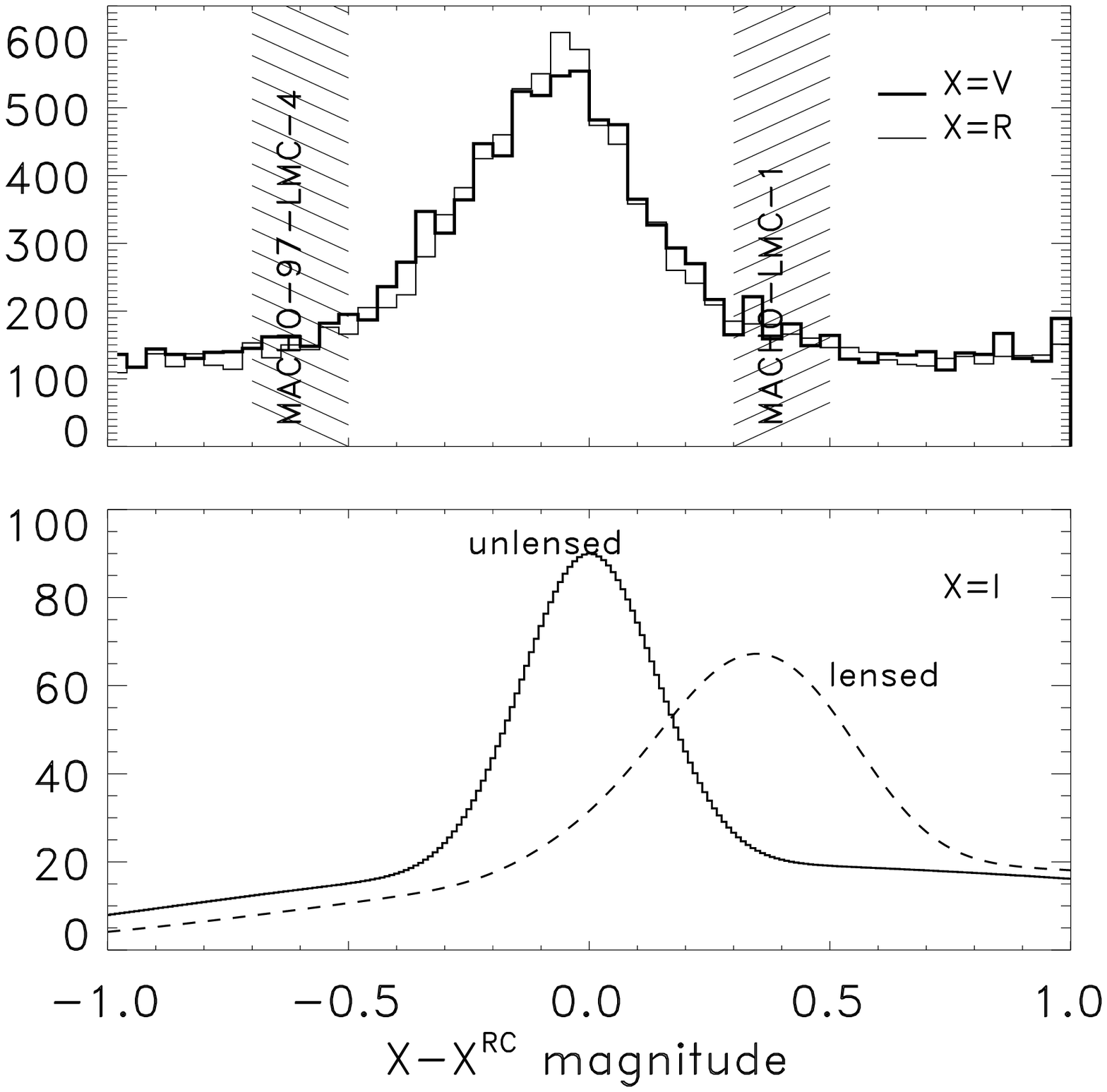}}
\caption{Luminosity functions of LMC stars in the MACHO $V$ and $R$ and OGLE $I$ bands. 
The plotted distributions of $X-X^{\rm RC}$,
the offset magnitude in the $X$ band from the red clump peak, should be insensitive to
any changes of zero point due to calibrations of 
the LMC distance modulus, foreground reddening and
the intrinsic luminosity of the red clump as well as changes of passbands, 
although the clump is narrower in the $I$ band.  
Overplotted in the upper panel 
are the event MACHO-LMC-1 and the alert MACHO-97-LMC-4 in the MACHO $V$ and $R$ bands.
Overplotted in the lower panel is the predicted $I$ band luminosity distribution of 
lensed stars (dashed curve), which is
clearly shifted to the fainter magnitude from the distribution of the unlensed stars
(solid curve).  The prediction assumes 90\% of the events involve a source
at a larger distance modulus than the LMC (by $0.2 - 0.5~{\rm mag}$ with equal probability).
}
\label{f1.eps}
\end{figure}

\label{lastpage}
\end{document}